\documentclass[english]{article}

\usepackage{geometry} 
\usepackage{graphicx}
\usepackage{amssymb}
\usepackage{epstopdf}

\usepackage{babel}
\usepackage{inputenc}
\usepackage{epsfig}

\newcommand{\Qto}       {Q_{\mathrm {t_{out}}}}
\newcommand{\Qts}       {Q_{\mathrm {t_{side}}}}
\newcommand{\Ql}        {Q_{\mathrm \ell}}
\newcommand{\qz}        {q_0}
\newcommand{\qt}        {q_{\mathrm t}}
\newcommand{\ql}        {q_{\mathrm \ell}}
\newcommand{\kt}        {k_{\mathrm t}}
\newcommand{\Rto}       {R_{\mathrm {t_{out}}}}
\newcommand{\Rts}       {R_{\mathrm {t_{side}}}}
\newcommand{\Rlong}     {R_{\mathrm {long}}}
\newcommand{\Rlongto}   {R_{\mathrm {long,t_{out}}}}
\newcommand{\Rz}        {R_0}
\newcommand{\Rt}        {R_{\mathrm t}}
\newcommand{\Rl}        {R_{\mathrm \ell}}
\newcommand{\eto}       {\epsilon_{\mathrm {t_{out}}}}
\newcommand{\ets}       {\epsilon_{\mathrm {t_{side}}}}
\newcommand{\elong}     {\epsilon_{\mathrm {long}}}
\newcommand{\dz}        {\delta_0}
\newcommand{\dt}        {\delta_{\mathrm t}}
\newcommand{\dl}        {\delta_{\mathrm \ell}}
\newcommand{\YYK}       {Y_{\mathrm {YK}}}

\begin{document}

\begin{center}
{\Large {\bf Bose-Einstein Correlations in Multihadron Events at LEP}}

\vspace{0.8cm}

\normalsize{
C. Ciocca, M. Cuffiani, G. Giacomelli \\
Dip. di Fisica of the University of Bologna and INFN Sezione di Bologna, 
I-40127 Bologna, Italy;\\ 
ciocca@bo.infn.it, cuffiani@bo.infn.it, giacomelli@bo.infn.it \\ } 

\par~\par

{\bf Invited paper at the ``Ninth Workshop on Non Perturbative QCD''\par
Institut d'Astrophysique de Paris, Paris, France, 4-8 June 2007}

\par~\par

\end{center}


{\bf Abstract.} {\normalsize
Bose-Einstein correlations in pairs of identical particles were analyzed in 
$e^+ e^-$ multihadron annihilations at $\sim$91.2 GeV at LEP. The first studies involved 
identical charged pions and the emitting source size was determined. 
Then the study of charged kaons suggested that 
the radius depends on the mass of the emitted particles. Subsequenty the 
dependence of the source radius on the event 
multiplicity was analyzed. The study of the correlations in neutral pions and 
neutral kaons extended these concepts to neutral particles. The shape of the 
source was analyzed in 3 dimensions and was found not to be 
spherically symmetric. In recent studies at LEP the correlations 
were analyzed in intervals of the average pair transverse momentum and of the pair 
rapidity to study the correlations between the pion production points and their 
momenta (position-momentum correlations). The latest $e^+ e^-$ data are consistent with 
an expanding source.}

\section{Introduction}
Bose-Einstein Correlations (BECs) are a quantum machanical phenomenon which 
manifests in final multihadron states as an enhanced probability for identical 
bosons to be emitted with small relative four momentum Q, compared with non 
identical bosons under similar kinematic conditions \cite{brown, gold, cuffiani}. 
From the measured effect it is possible to determine the space 
time dimensions of the boson-emitting source. 
The BEC effect  arises from the ambiguity of path between 
sources and detectors and the requirement to symmetrise the wave function of two 
or more identical bosons. \par

In 1954 the radioastronomers R. Hanbury-Brown and R. Q. Twiss proposed a new 
interferometry technique to measure the angular dimension of a star. 
It required to measure the mixed intensities in two 
radiotelescopes; the dependence of the correlation on the distance between them 
yielded the angular diameter of the astronomical source 
\cite{brown}. 
G. Goldhaber et al applied the same principle in particle physics, 
in $\bar p p$ annihilations into two identical charged pions, obtaining the radius of the emitting source, $\sim$1 fm \cite{gold}. \par

The first LEP analyses  on BECs concerned identical charged pions, assuming a spherical emission source and yielded the size of the source (R$\sim$1 fm) and the 
chaoticity parameter \cite{cuffiani, acton}. Neutral $\pi^{\circ}$ were then 
considered \cite{abbiendi}.
Then the study was extended to neutral and charged kaons in order to determine if the source radius depends on the mass of the emitted particles \cite{kk, alexander}.

\begin{figure}
 \begin{minipage}[b]{8.5cm}
   \centering
   \includegraphics[width=8cm]{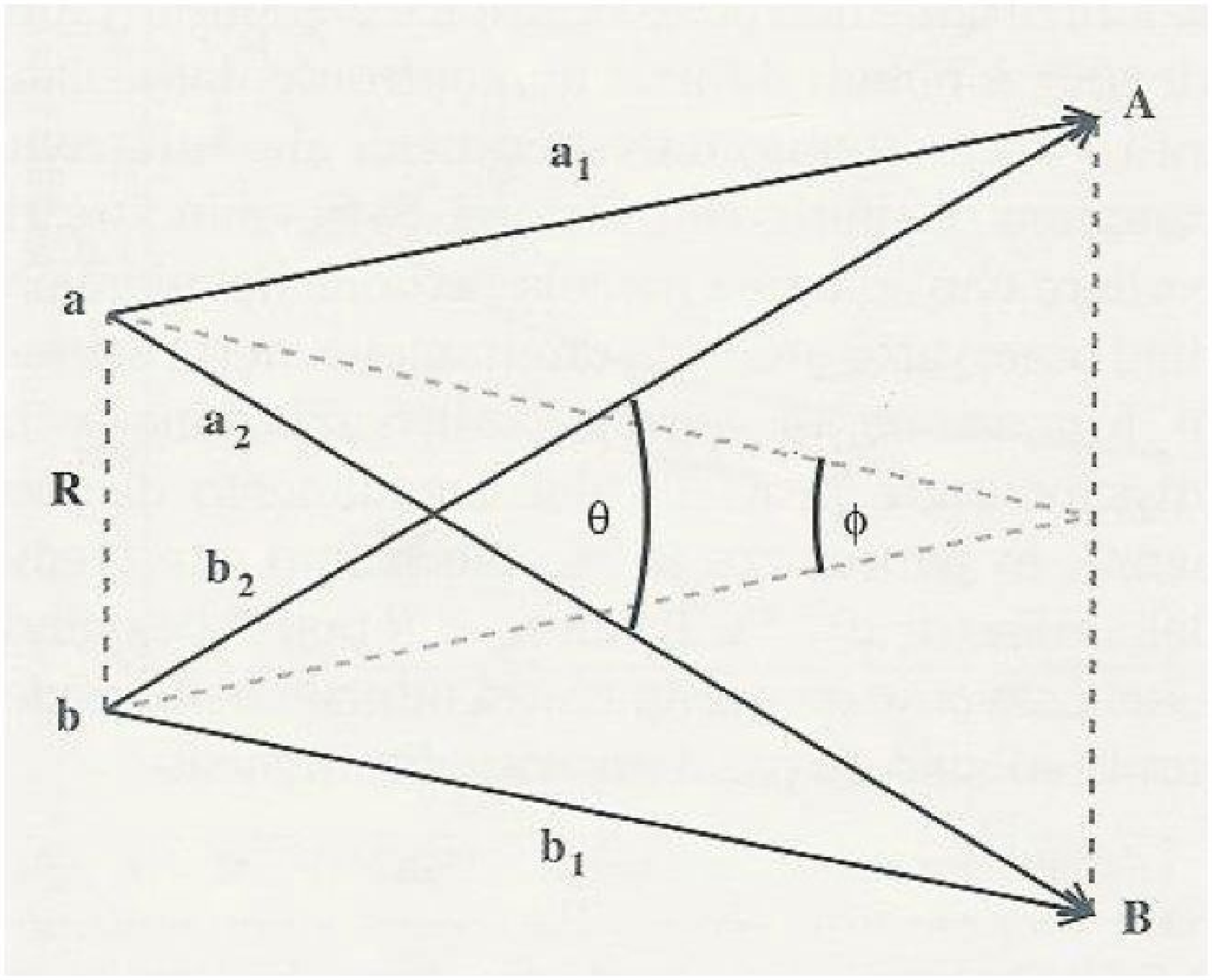}
\begin{quote}   
\caption{\small Scheme of a measurement of BECs. a, b are two sources 
separated by a distance R; A, B are two detectors separated by a distance 
L. The 
emitted particles go from sources to detectors as a$\rightarrow$A, 
b$\rightarrow$B or as a$\rightarrow$B, b$\rightarrow$A. In astronomy L$<<$R, 
in particle physics L$>>$R.}
\label{scheme}
\end{quote} 
 \end{minipage}
 \begin{minipage}[b]{7.5cm}
  \centering
   \includegraphics[width=8.5cm]{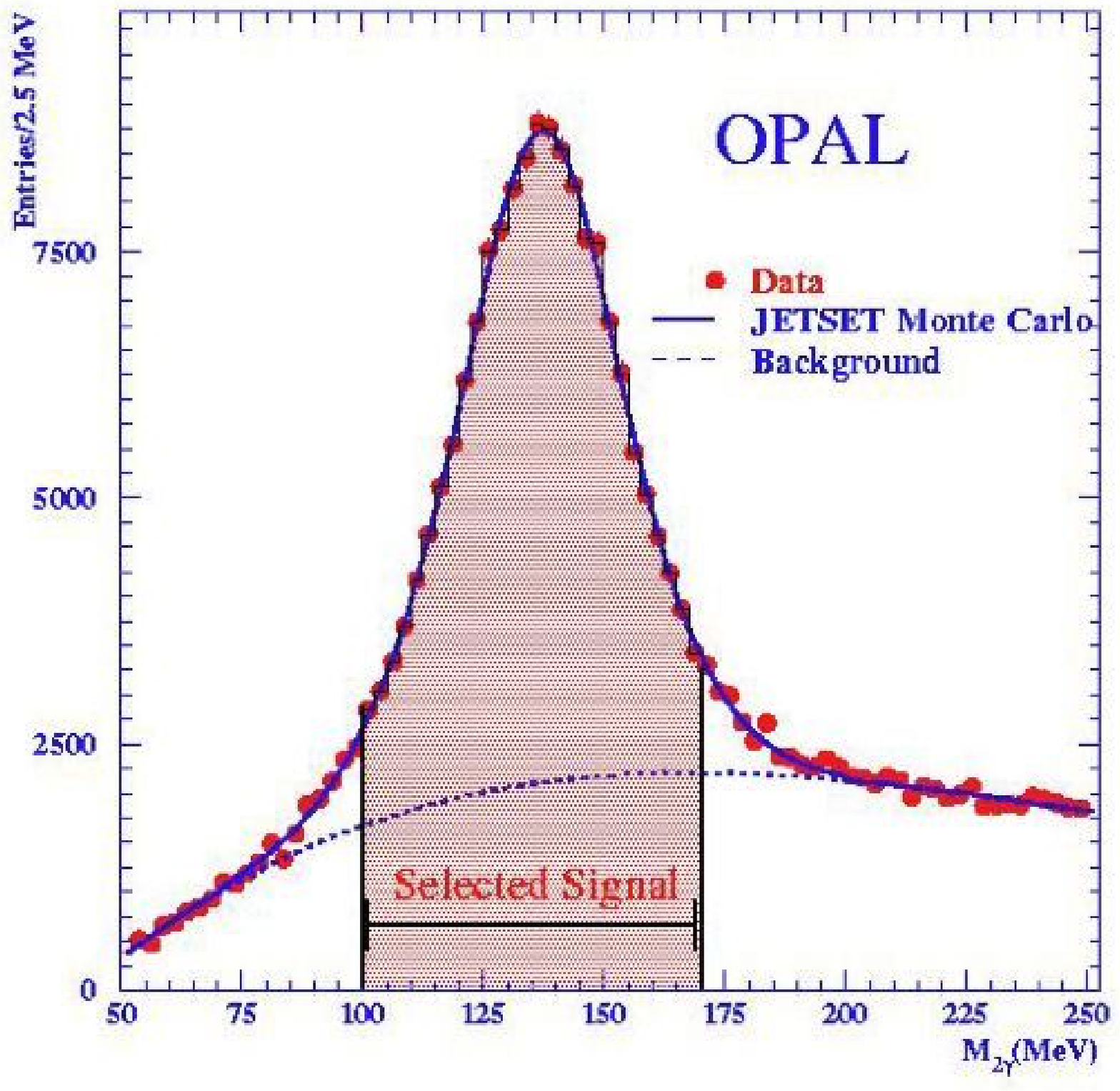}
\begin{quote}   
\caption{\small Distribution of two-photon invariant mass, $M_{2 \gamma}$. The smooth curves are the total Monte Carlo expectation (solid line) and the background expectation (dashed line). The shaded region is the selected window for the $\pi^0$ signal. }
\label{signal}
\end{quote}
 \end{minipage}
\end{figure}

Further analyses were performed to establish if the emitting source radius depends on the particle multiplicity \cite{gg}. Other studies involved the search for BE correlations in multipions \cite{ackerstaff}.\par

BECs were studied in two and three dimensions and one discovered that the emitting source is not spherical \cite{multipions}. Many studies were made for WW correlations \cite{GA} and also in $\nu$ interactions \cite{allasia}. 
Finally they involved the study of expanding sources and trials to determine the emission time \cite{momentum}.\par

We shall make a brief survey of these studies, concentrating finally on the very recent works which prove that even in $e^+ e^-$ collisions one has expanding sources.

\section{Experimental Procedure}
A detailed description of the OPAL experiment may be found  in ref \cite{ahmet}. The most important subdetector for BEC studies is the Central Tracking 
Detector, the Jet Chamber. For $\pi^0$ studies we needed also the barrel electromagnetic calorimeter. 
A sample of 4.3 million multihadronic events from $Z^0$ decays were used. 
A set of quality cuts was applied and one used cuts specific for BEC studies.\par

First, the event thrust axis was computed, using tracks with a
minimum of 20 hits in the jet chamber, a minimum transverse momentum of 
150 MeV and a maximum momentum of 65 GeV.
Clusters in the electromagnetic calorimeter were used for energies
exceeding 100 MeV in the barrel or 200 MeV in the endcaps.
Only events well contained in the detector were accepted,
requiring $|{\rm cos}\theta_{\mathrm {thrust}}|<0.9$,
where $\theta_{\mathrm {thrust}}$ is
the polar angle of the thrust axis with respect to the beam axis. Tracks were required to have a maximum momentum of 40 GeV and to
originate from the interaction vertex.
Electron-positron pairs from photon conversions were rejected.
The selected events contained a minimum of five 
tracks and were reasonably balanced in charge, i.e. 
$|n_{\mathrm {ch}}^{+}-n_{\mathrm {ch}}^{-}|/(n_{\mathrm {ch}}^{+}+
n_{\mathrm {ch}}^{-}) \leq 0.4$,
where $n_{\mathrm {ch}}^{+}$ and $n_{\mathrm {ch}}^{-}$ are the number
of positive and negative charge tracks, respectively.
About 3.7 million events were left after all cuts.
All charged particle tracks that passed the selections were used,
the pion purity being approximately 90\%. \par

Since in multihadron events more than 90\% of the measured tracks are charged pions, the study of BEC for like-sign charged pion pairs was usually performed without proper particle identification and without purity correction. This choice 
introduces a small error in the chaoticity parameter $\lambda$ and in the 
radius R of the emitting region. However some analyses were performed with 
properly identified pions. That required some effective cuts on the fraction of the global solid angle acceptance.\par

For $\pi^0 \pi^0$, $K^{\pm}K^{\pm}$ and $K^0 K^0$ correlations, 
particle identification was necessary \cite{abbiendi, kk}. 
Fig. 2 shows the distribution of the two-photon invariant mass and the $\pi^0$ event 
selection.

\section{BECs from a static source}
{\bf BECs in one dimension.} The measured BEC function is defined as the ratio 
$~C(Q)=\rho(Q)/\rho_0(Q)~$, were $Q$ is a Lorenz-invariant variable expressed in 
terms of the two pion four momenta $p_1$ and $p_2$ as 
$Q^2$=-($p_1$ - $p_2$)$^2$, $\rho(Q) = (1/N) dN/dQ$ is the measured $Q$ 
distribution of the two pions and $\rho_0 (Q)$ 
is a reference distribution which should contain all the correlations included 
in $\rho(Q)$, except BECs. For the determination of $\rho_0(Q)$, 
different methods were used: for identical $\pi^+ \pi^+$ and $\pi^- \pi^-$ one used 
the $\pi^+ \pi^-$ sample, but also the event mixing reference sample, 
where pion pairs are formed from pions belonging to different events; 
also a Monte Carlo (MC) reference sample without BECs was used. \par

The correlation distribution $C(Q)$ was parametrised using the Fourier 
transform of the expression for a static sphere of emitters with a Gaussian density:
\begin{equation}
C(Q) =N(1+ \lambda exp(-R^2 Q^2))(1 + \delta Q + \epsilon Q^2).
\end{equation}

\noindent $\lambda$  is the chaoticity parameter, $R$ is the radius of the source, 
and $N$ a normalization factor. The empirical term $(1 + \delta Q + \epsilon Q^2)$ 
accounts for the behaviour of the correlation function at high $Q$ due to 
any remaining long-range correlation.
The largest difference among results from different experiments lies in the choice 
of the reference sample: the statistical errors on $R$ is small, but the systematic 
uncertainty is large. \par

Fig 3 shows a typical distribution of $C(Q)$ versus $Q$; it is relative to 
a three dimensional analysis, but the observed features are typical of all BECs: notice 
the BEC peak at low $Q$ and the tail at large $Q$; the solid line is a fit 
to eq. 1, excluding the $Q$-intervals indicated in the figure, which contain 
effects from known hadron resonances.\par

\begin{figure}
 \begin{minipage}[b]{8cm}
   \centering
   \includegraphics[width=6cm]{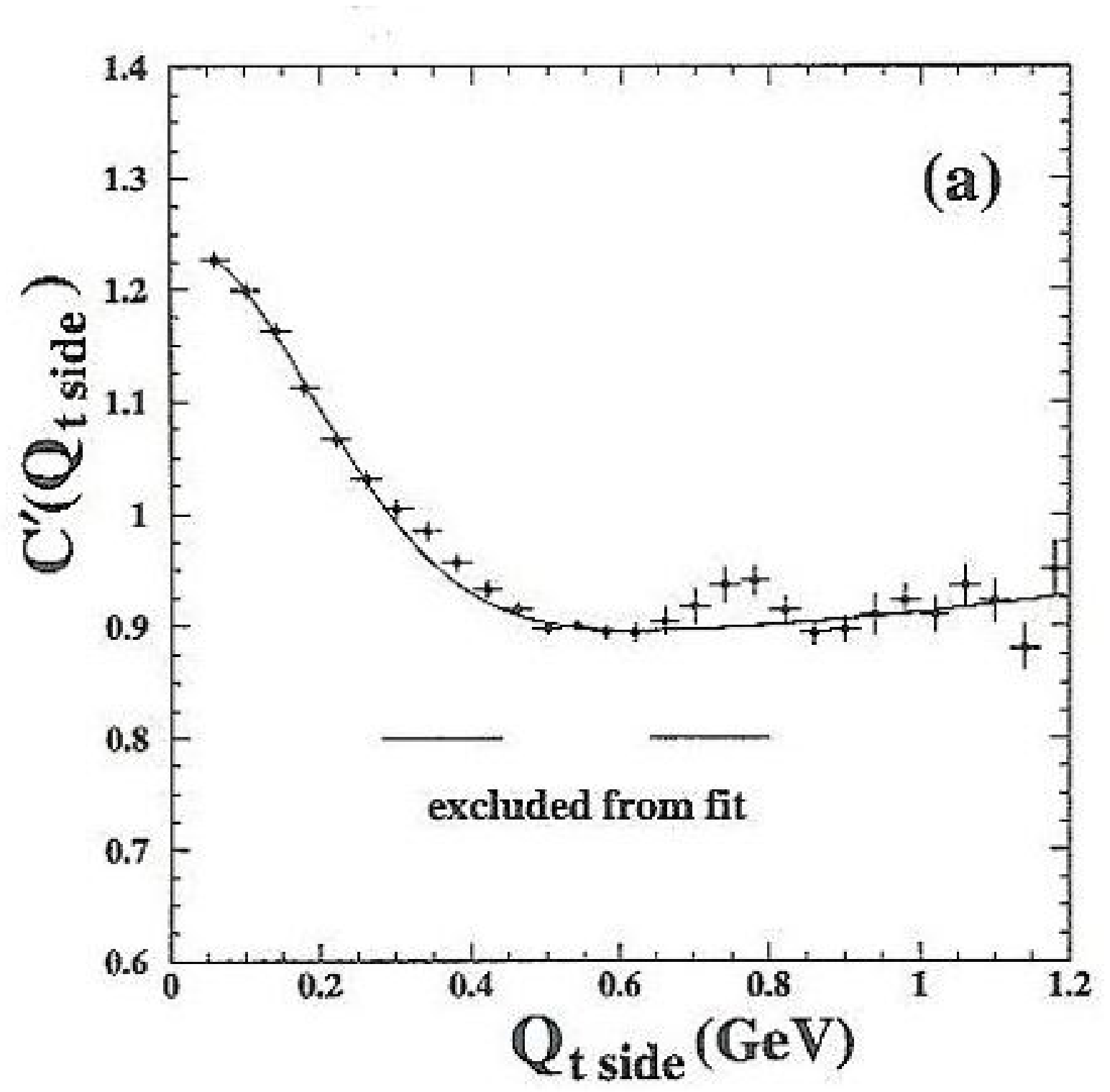}
\begin{quote}
\caption{\small The BEC distribution $C'(Q)$ for charged pions vs $\Qts$. 
The smooth solid curve is the fitted correlation function (the excluded 
regions contain effects from known hadron resonances). }
\label{fig3}
\end{quote}   
 \end{minipage}
 \begin{minipage}[b]{8.5cm}
  \centering
   \includegraphics[width=8.5cm]{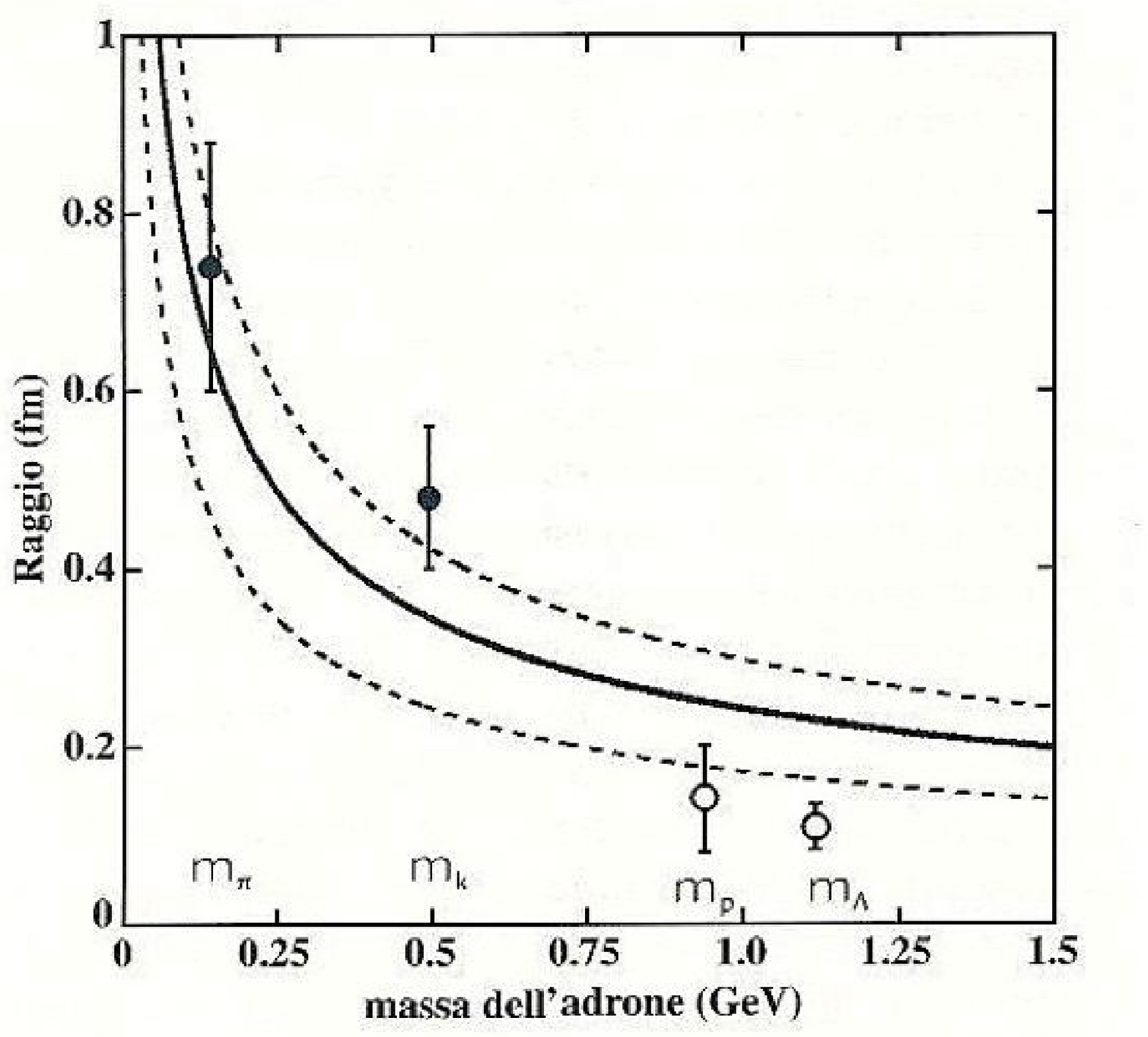}
   \begin{quote}
\caption{\small Radius of the emitting region for BECs of 2 identical bosons and 
for FDCs of 2 identical baryons produced in $e^+ e^-$ collisions at LEP. }
\label{fig4}
\end{quote}
 \end{minipage}
\end{figure}

The same analysis was repeated for $KK$ BECs. A similar analysis was performed on 
Fermi Dirac Correlations (FDCs) for identical fermions: in this case there is no 
peak at small values of $Q$, but a dip. The analysis gives the radius of the 
emitting regions as shown in Fig. 4. Note that there probably is a decrease of 
$R$ with increasing mass of the emitted identical particles.\par

Fig. 5a shows the variation of the emitting radius with the charged multiplicity of the event \cite{gg}: there is an increase of about 10\% of the radius when the 
multiplicity increases from 10 to 40 charged hadrons in 
the final state. This may be related to the number of hadron jets: one has 
$R_{4jets} > R_{3jets} > R_{2jets}$. Notice that there is a corresponding 
decrease of the chaoticity parameter, Fig. 5b.\par

\begin{figure}
 \begin{minipage}[b]{8.5cm}
   \centering
   \includegraphics[width=10.5cm]{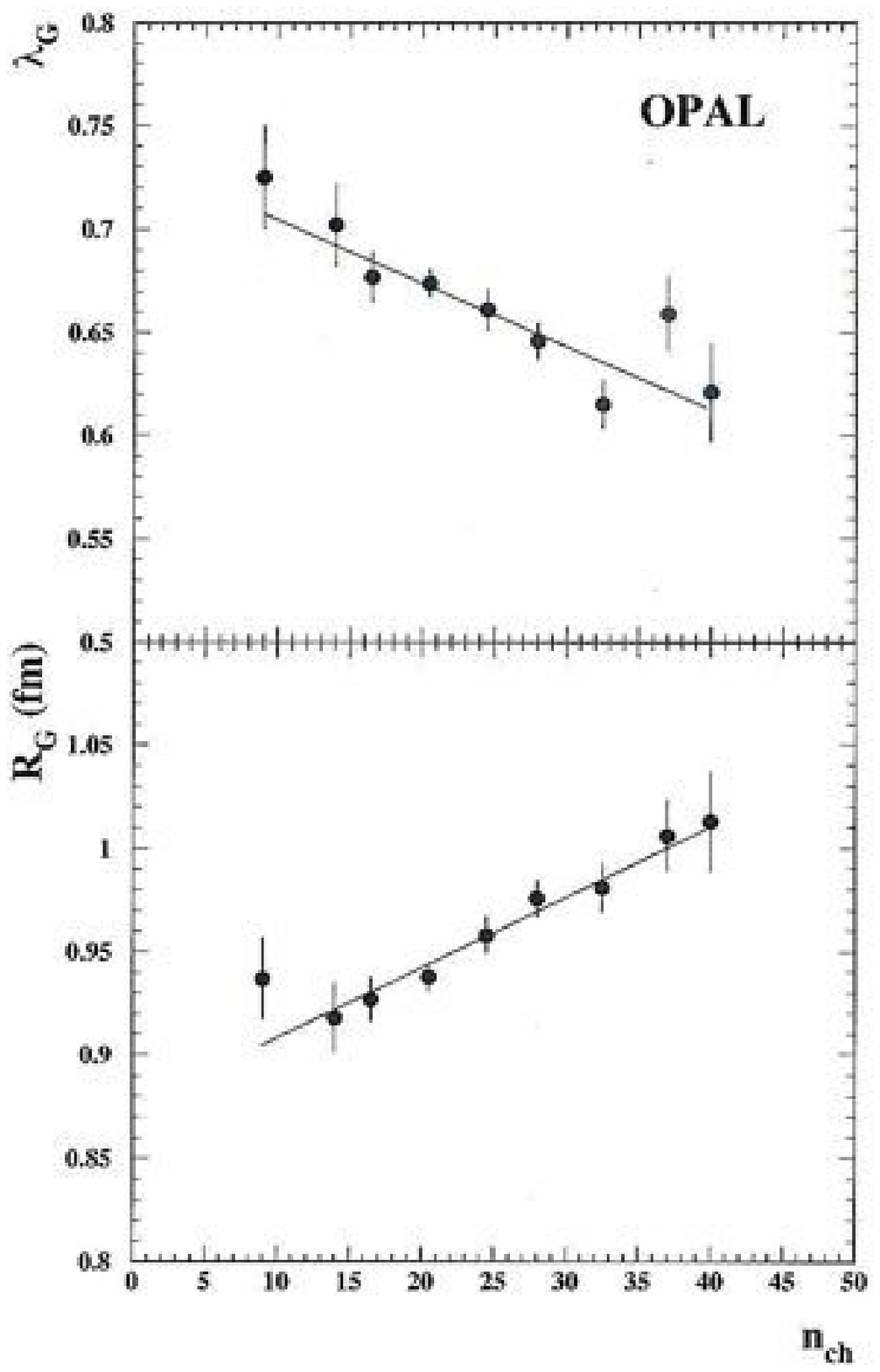}
\begin{quote}
\caption{\small a) Increase of the emitting radius with increasing event 
multiplicity and b) decrease of $\lambda$. }
\label{fig5}
\end{quote}   
 \end{minipage}
 \begin{minipage}[b]{8.5cm}
 \centering
   \includegraphics[width=9cm]{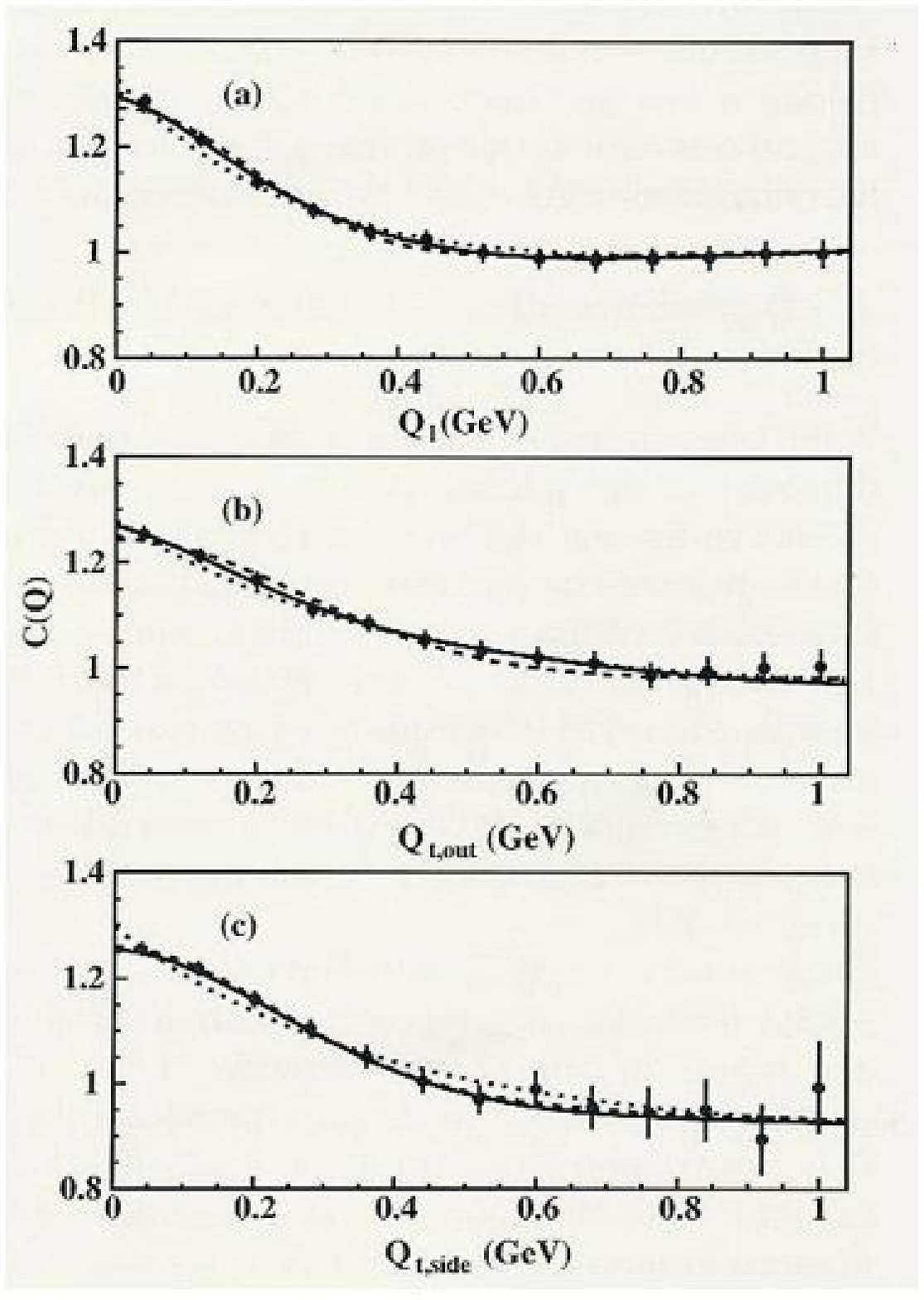}
   \begin{quote}
\caption{\small The distributions in $\Qto$ and $\Qts$ are broader than in $\Ql$: 
thus $\Rl > \Rts \sim \Rto$.  }
\label{fig7}
\end{quote}
 \end{minipage}
\end{figure}

In ref. \cite{multipions} it was found that there are 3$\pi$ BECs, that is 
after removing the effect of 2$\pi$ correlations on the 3$\pi$ sample. 
The present situation is consistent with the relation 

\begin{equation}
R_{3\pi} = R_{2\pi} / \sqrt{2}
\end{equation}

In ref. \cite{ackerstaff} it was found that there are true multiparticle correlations up to 5$\pi$.\par

\begin{figure}
 \begin{minipage}[b]{8.5cm}
   \centering
   \includegraphics[width=9cm]{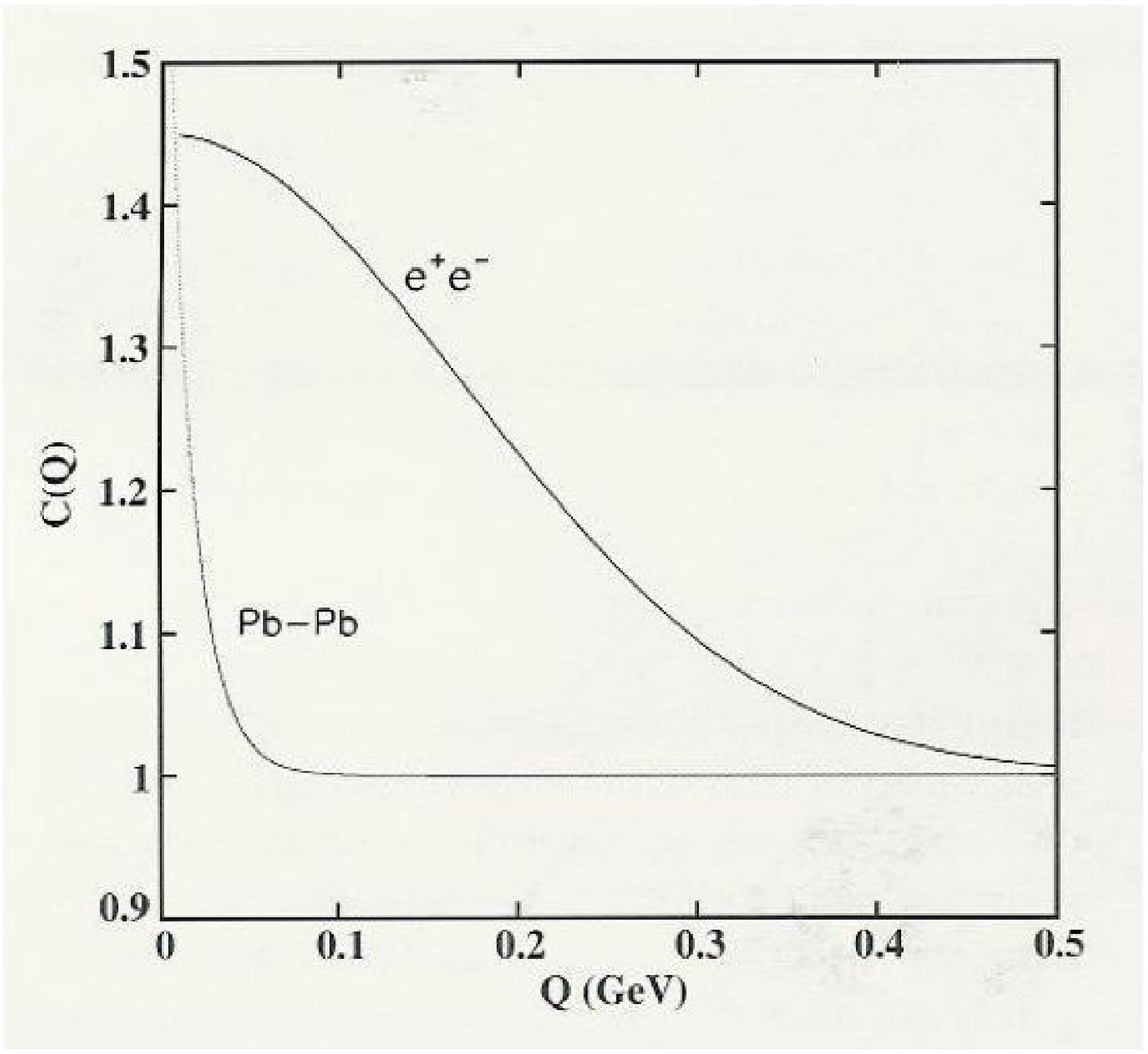}
\begin{quote}   
\caption{\small Comparison of the BE Correlation functions for PbPb and 
$e^+ e^-$ collisions. }
\label{fig6}
\end{quote}
 \end{minipage}
 \begin{minipage}[b]{8.5cm}
  \centering
   \includegraphics[width=6.5cm]{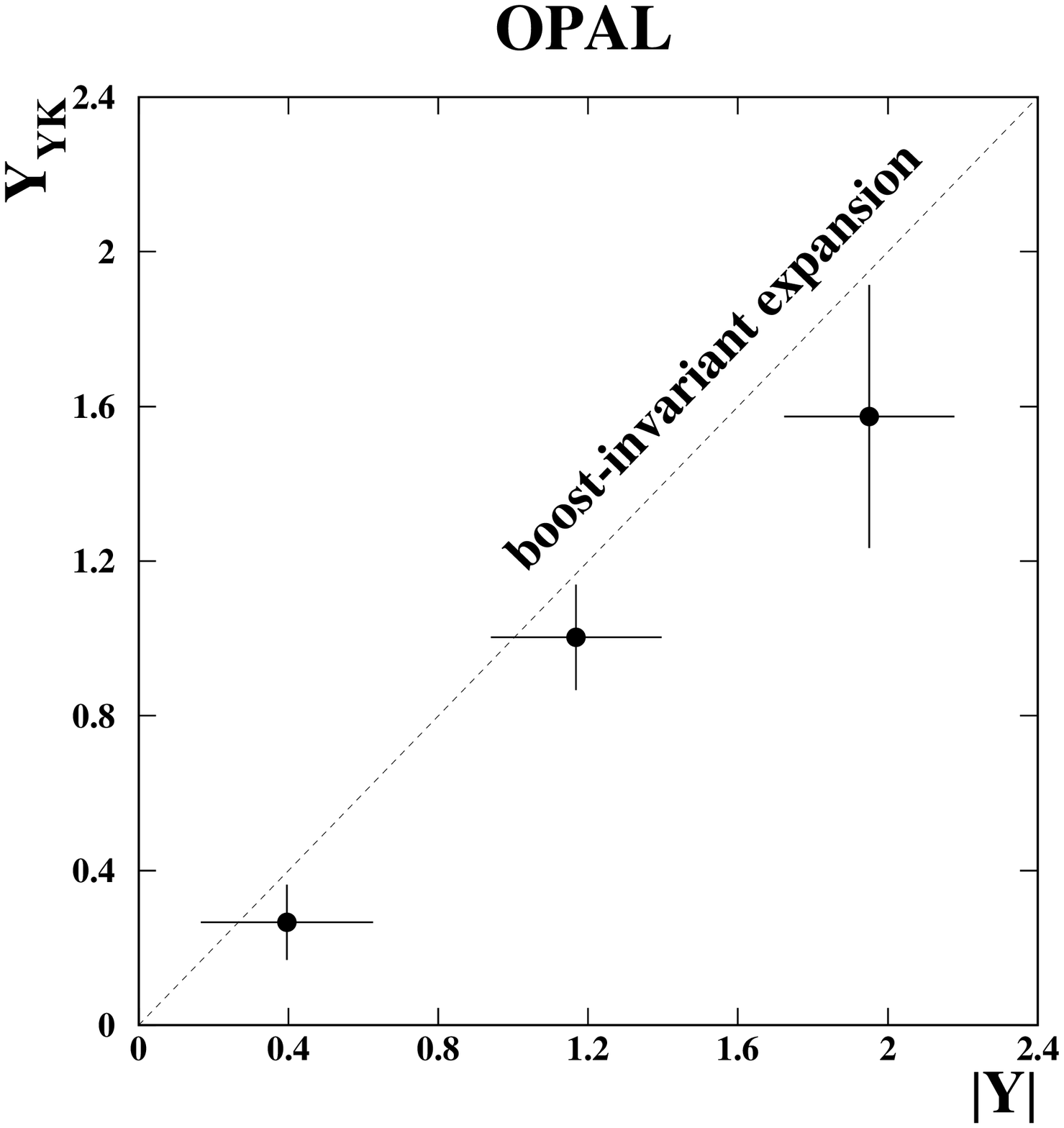}
   \begin{quote}
\caption{\small $\YYK$ vs pion pair rapidity $Y$. Vertical bars 
include statistical and systematic errors. $\YYK$=$Y$ corresponds 
to a source expanding boost-invariantly.}
\label{fig:yy}
\end{quote}
 \end{minipage}
\end{figure}

\noindent {\bf BECs in two and three dimensions.} 
Multidimensional static analyses were performed in 3 dimensions using the 
Longitudinal Center of Mass System (LCMS): the sum of the impulses of the emitted 
$q \bar q$ pair lies in the plane perpendicular to the event axis, defined by the $q \bar q$ direction. 
The components of the  3-dimensional distribution in the longitudinal, 
out and side projections indicate that the last ones are larger, see Fig. 6. Thus the longitudinal radius is about 20\% larger than the transverse radius: the 
emitting source is ellissoidical, elongated in the $q \bar q$ direction.\par

\noindent {\bf Comparison of BECs in $e^+ e^-$ and Nucleus-Nucleus collisions.}
Fig. 7 shows the BEC functions in $e^+ e^- \rightarrow$ hadrons  and 
 Pb Pb $\rightarrow$ hadrons: note how much narrower is the 
distribution in Pb Pb collisions: the distribution yields a radius 
$R \simeq 6-7$ fm for the emissions of pion pairs.

\section{Expanding sources}
BECs have been analyzed in Nucleus-Nucleus collisions in order to find evidence for expanding sources due to the formation of a quark-gluon deconfined plasma \cite{aggraval, adams}. 
Expanding sources may arise in $e^+ e^-$ collisions because of string fragmentation \cite{geiger}. In order to study BECs in non static, expanding sources we analyze the correlation functions

\begin{equation}
C'=\frac{C^{\rm DATA}}{C^{\rm MC}}=
\frac{N^{\rm DATA}_{\rm like}/N^{\rm DATA}_{\rm unlike}}
{N^{\rm MC}_{\rm like}/N^{\rm MC}_{\rm unlike}},
\end{equation}

in bins of the average pair four-momentum with respect to the event thrust direction

\begin{equation}
k_t =(\vec{p}_{{\mathrm t},1} + \vec{p}_{{\mathrm t},2}) 
\end{equation}

and of the pair rapidity:

\begin{equation}
|Y| = \frac{1}{2} \ln \left[
{\frac{(E_1+E_2)+(p_{{\mathrm \ell},1}+p_{{\mathrm \ell},2})}
{(E_1+E_2)-(p_{{\mathrm \ell},1}+p_{{\mathrm \ell},2})}}\right]
\end{equation}

The experimental distributions in $dN / d|Y|$ and $dN /dk_t$ 
are in good agreement with the distributions from the Jetset Monte Carlo. 
The dependences of $C$ and $C'$ on $K$ were studied in three bins of $|Y|$
($0.0\leq|Y|<0.8$, $0.8\leq|Y|<1.6$, $1.6\leq|Y|<2.4$) and five bins of 
$\kt$ ($0.1\leq\kt<0.2$ GeV, $0.2\leq\kt<0.3$, $0.3\leq\kt<0.4$, 
$0.4\leq\kt<0.5$ and $0.5\leq\kt<0.6$ GeV).\par

\begin{figure}[!h]
\centerline{\includegraphics[width=10cm]{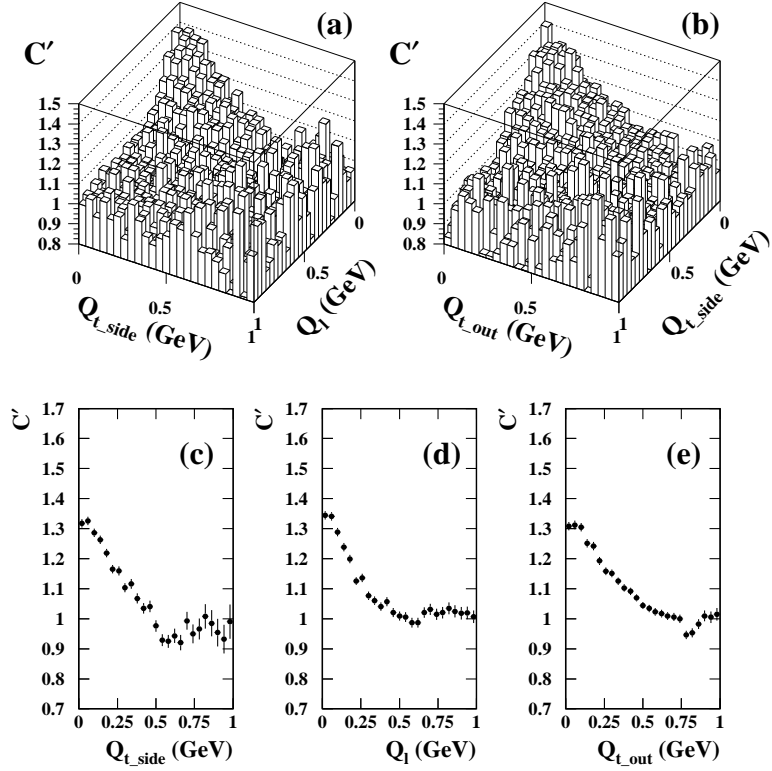}}
\begin{quote}
\caption{\small Two-dimensional (a), (b) and one-dimensional (c), (d) and (e) 
projections of the correlation function $C'(\Ql,\Qts,\Qto)$ 
for 0.8 $\leq |Y| <$ 1.6 and 0.3 $\leq \kt <$ 0.4 GeV. $\Qto <$ 0.2 GeV in 
(a), $\Ql <$ 0.2 GeV in (b).  In (c), (d), (e) the projections at low values 
($<$0.2 GeV) of the other variables.}
\label{proBP_2}
\end{quote}
\end{figure}

Two-dimensional projections of the correlation function 
$C'(\Ql,\Qts,\Qto)$ for a single bin of $|Y|$ and $\kt$ are shown
in Fig.~\ref{proBP_2}a, b. BEC peaks are visible at low $\Ql,\Qts,\Qto$. 

\par

To extract the spatial and temporal extensions of the pion source from
the experimental correlation functions,
the Bertsch-Pratt (BP)
\vspace{-0.2cm}

\begin{equation}
C'(\Ql,\Qts,\Qto) = N (1 + \lambda {\rm e}^{-(\Ql^2 \Rlong^2 +
\Qts^2 \Rts^2 +
\Qto^2 \Rto^2 +
2 \Ql \Qto \Rlongto^2
)})
F(\Ql,\Qts,\Qto)
\end{equation}
and the Yano-Koonin (YK)
\begin{equation}
C'(\qt,\ql,\qz) =
N (1 + \lambda {\rm e}^{-(\qt^2 \Rt^2 +
\gamma^2(\ql - v \qz)^2 \Rl^2 +
\gamma^2(\qz - v \ql)^2 \Rz^2
)})
F(\qt,\ql,\qz)
\end{equation}

\noindent parameterizations were fitted to the measured correlation
functions in intervals of $\kt$ and $|Y|$. 
In both parameterizations, $N$ is a normalization factor, 
$\lambda$ is the degree of incoherence of the pion sources, related to the fraction of pairs that interfere.
The parameters $N$ and $\lambda$, whose product determines the size of the
BEC peak, are significantly (anti)correlated. The two functions 
$F(\Ql,\Qts,\Qto) =
(1 + \elong\Ql + \ets\Qts + \eto\Qto)$
and $F(\qt,\ql,\qz) = (1 + \dt\qt + \dl\ql + \dz\qz)$,
where $\epsilon_{\mathrm i}$ and $\delta_{\mathrm i}$ are free
parameters, were introduced in Eq.~(6) and (7) to take into
account residual long-range two-particle correlations due to energy and
charge conservation. The interpretation of the other parameters in Eq~(6), is:\\
- $\Rts$ and $\Rlong$ are the transverse and longitudinal 
radii in the longitudinal rest frame of the pair;\\
- $\Rto$ and the cross-term $\Rlongto$ are a
combination of both spatial and temporal extentions of the source.
The difference ($\Rto^2 - \Rts^2$) is proportional to the duration of
the particle emission process, and $\Rlongto$ to the source
velocity with respect to the pair rest frame.\par

In the YK frame Eq.~(7), where $\gamma=1/\sqrt{1-v^2}$, the free
parameters are interpreted as follows:\\
- $v$ is the longitudinal velocity, in units of $c$,
of the source element in the CMS frame;\\
- $\Rz$ measures the time interval, times $c$, during which
particles are emitted, in the rest frame of the emitter (source element).
The limited phase-space available limits the analysis for $\Rz^2$ ;\\
- $\Rt$ and $\Rl$ are the transverse and longitudinal radii, in the rest frame of the
emitter.\par

The parameters $\Rz$, $\Rt$ and $\Rl$ are
evaluated in the rest frame of the source element. 
The two parameterizations are not independent, so that a
comparison between the BP and YK fits is an important test.
In the YK picture the source velocity $v$ of each element does not depend on 
$k_t$, while it is correlated with the pair rapidity Y. $Y_{YK}$ measures the 
rapidity of the source element with respect to the c.m frame 
$Y_{YK} = \frac{1}{2} \ln \left[ (1+v) / (1-v) \right]$. 
A non expanding source corresponds to $Y_{YK} \simeq 0$ for any Y. For a 
longitudinally boost invariant source (for which the velocity of each element 
is $v = z/t$, where $t$ is the time elapsed since the collision and $z$ is the 
longitudinal coordinate of the element) the correlation $Y_{YK} =Y$ 
is expected as in Fig. 8. \par

The following relations hold between the BP and YK parameters:
\begin{eqnarray}
\Rts^2 & = & \Rt^2 \\
\Rlong^2 & = &
\gamma_{\mathrm {LCMS}}^2(\Rl^2+\beta_{\mathrm {LCMS}}^2 \Rz^2) \\
(\Rto^2-\Rts^2) & = & \beta_{\mathrm t}^2
\gamma_{\mathrm {LCMS}}^2(\Rz^2+\beta_{\mathrm {LCMS}}^2\Rl^2).
\end{eqnarray}
$\beta_{\mathrm {LCMS}}$ is the velocity of the
source element
in the LCMS, i.e. with respect to the pair longitudinal rest frame;
$\gamma_{\mathrm {LCMS}}=1/\sqrt{1-\beta_{\mathrm {LCMS}}^2}$. 
For a boost-invariant source, $\beta_{\mathrm {LCMS}}=0$: (9)
and (10) become:
\begin{eqnarray}
\Rlong^2 & \simeq & \Rl^2 \\
(\Rto^2-\Rts^2) & \simeq & \beta_{\mathrm t}^2 \Rz^2.
\end{eqnarray}
\par
In Fig. 10 the best-fit BP and YK parameters are compared: \\
- The longitudinal parameter $\Rlong^2$ is larger than
$\Rl^2$ in all rapidity intervals, Fig. 9(a),(d) and (g). $\Rlong^2 > \Rl^2$
corresponds to $\beta_{\mathrm {LCMS}}$ $>$0, i. e. to a pion source 
whose expansion is not exactly boost-invariant.\\
- The equality of the transverse parameters, $\Rts^2$ $\Rt^2$, is 
confirmed; there may be deviations at low $\kt$. \\
- $R^2_0$ and $(\Rto^2-\Rts^2)$ are essentially equal to zero, suggesting that the present technique does not allow to mesure the duration of the emission process.

\begin{figure}[!h]
\centerline{\includegraphics[width=14cm]{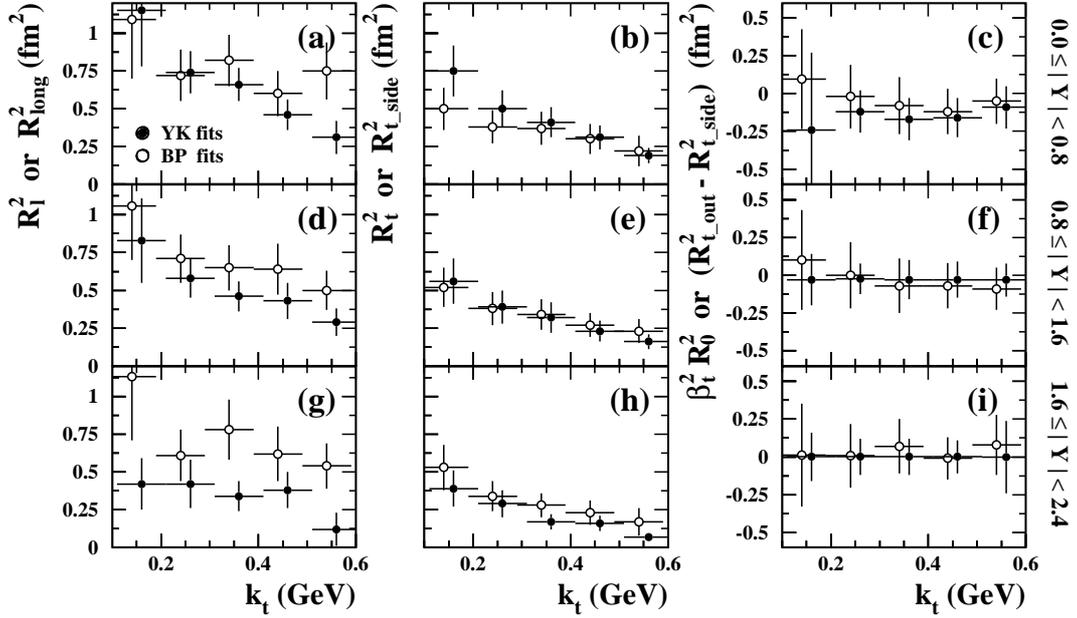}}
\begin{quote}
\caption{\small BP and YK fits. (a)(d)(g) The best-fit longitudinal 
radius $\Rlong^2$ in the BP frame (open dots) compared 
with the 
YK $\Rl^2$ (full dots). (b)(e)(h) The BP transverse correlation length 
$\Rts^2$ (open dots) compared with the YK $\Rt^2$ (full dots). (c)(f)(i) The difference of the BP transverse radii $(\Rto^2-\Rts^2)$ (open dots) compared with the YK time parameter $\Rz^2$ times $\beta_{\mathrm t}^2$ (full dots). }
\label{all9par_bp}
\end{quote}
\end{figure}
\par

\section{Conclusions}
We have first summarized the results obtained on BECs in 
$e^+ e^-$ collisions at the $Z^0$ peak assuming a static source. Then we presented 
an analysis in bins of the average 4-momentum of the pair. Based on this, 
the dynamic features of the pion emitting source were investigated in the YK 
and BP formalisms.\par
 
The transverse and longitudinal radii of the pion sources
decrease for increasing
$\kt$, indicating the presence of correlations between the particle
production points and their momenta.
The YK rapidity scales with the pair rapidity,
in agreement with a nearly boost-invariant expansion of the pion source.
Phase space limitation did not allow the measurement of the
duration of the particle emission process.
\par
Similar results have been observed in more complex systems, such as the
pion sources created in pp and heavy-ion collisions, which are now
complemented with measurements in the simpler hadronic system formed
in $e^+ e^-$ annihilations. The unexplained similarities between BECs in 
different reactions might indicate a present limitation of our understanding of these correlations \cite{csorgo}.\\

{\bf Acknowledgements}. We thank all the members of the OPAL Collaboration, in particular the Bologna members. We acknowledge the contribution of Ms. Anastasia Casoni.

\end{document}